\documentclass[twocolumn]{aastex62}

\graphicspath{{./}{figures/}}

\received{September 15, 2019}
\revised{November 1, 2019}
\accepted{November 7, 2019 in ApJS}

\shorttitle{Tracing the origin of density structures}
\shortauthors{Rouillard et al.}

\usepackage[utf8]{inputenc}

\begin{document}

\title{Relating streamer flows to density and magnetic structures at the \textit{Parker Solar Probe}}

\correspondingauthor{Alexis P. Rouillard}

\author{Alexis P. Rouillard}
\affiliation{IRAP, Universit\'e Toulouse III - Paul Sabatier,
CNRS, CNES, Toulouse, France}
\email{arouillard@irap.omp.eu}

\author{Athanasios Kouloumvakos}
\affiliation{IRAP, Universit\'e Toulouse III - Paul Sabatier,
CNRS, CNES, Toulouse, France}

\author[0000-0002-8164-5948]{Angelos Vourlidas}
\affiliation{Johns Hopkins University Applied Physics Laboratory, Laurel, MD 20723, USA}

\author[0000-0002-7077-930X]{Justin Kasper}
\affiliation{Climate and Space Sciences and Engineering, University of Michigan, Ann Arbor, MI 48109, USA}
\affiliation{Smithsonian Astrophysical Observatory, Cambridge, MA 02138 USA}

\author[0000-0002-1989-3596]{Stuart Bale}
\affiliation{Physics Department, University of California, Berkeley, CA 94720-7300, USA}
\affiliation{Space Sciences Laboratory, University of California, Berkeley, CA 94720-7450, USA}
\affiliation{The Blackett Laboratory, Imperial College London, London, SW7 2AZ, UK}
\affiliation{School of Physics and Astronomy, Queen Mary University of London, London E1 4NS, UK}

\author{Nour-Edine Raouafi}
\affiliation{Johns Hopkins University Applied Physics Laboratory, Laurel, MD 20723, USA}

\author[0000-0001-6807-8494]{Benoit Lavraud}
\affiliation{IRAP, Universit\'e Toulouse III - Paul Sabatier,
CNRS, CNES, Toulouse, France}

\author[0000-0001-9027-8249]{Russell A. Howard}
\affiliation{Naval Research Laboratory, Washington, DC, USA}

\author[0000-0001-8480-947X]{Guillermo Stenborg}
\affiliation{Naval Research Laboratory, Washington, DC, USA}

\author[0000-0002-7728-0085]{Michael Stevens}
\affiliation{Smithsonian Astrophysical Observatory, Cambridge, MA 02138 USA}

\author[0000-0002-1814-4673]{Nicolas Poirier}
\affiliation{IRAP, Universit\'e Toulouse III - Paul Sabatier,
CNRS, CNES, Toulouse, France}

\author{Jackie A. Davies}
\affiliation{RAL Space, STFC-Rutherford Appleton Laboratory, Didcot, United Kingdom}

\author[0000-0003-1377-6353]{Phillip Hess}
\affiliation{Naval Research Laboratory, Washington, DC, USA}

\author[0000-0003-1380-8722]{Aleida K. Higginson}
\affiliation{Johns Hopkins University Applied Physics Laboratory, Laurel, MD 20723, USA}

\author{Michael Lavarra}
\affiliation{IRAP, Universit\'e Toulouse III - Paul Sabatier,
CNRS, CNES, Toulouse, France}

\author[0000-0003-1692-1704]{Nicholeen M. Viall}
\affiliation{Heliophysics Division, NASA Goddard Space Flight Center, Greenbelt, MD  20771, USA}

\author[0000-0001-6095-2490]{Kelly Korreck}
\affiliation{Smithsonian Astrophysical Observatory, Cambridge, MA 02138 USA}

\author[0000-0001-8247-7168]{Rui F. Pinto}
\affiliation{IRAP, Universit\'e Toulouse III - Paul Sabatier,
CNRS, CNES, Toulouse, France}

\author[0000-0001-8956-2824]{L\'ea Griton}
\affiliation{IRAP, Universit\'e Toulouse III - Paul Sabatier,
CNRS, CNES, Toulouse, France}

\author[0000-0002-2916-3837]{Victor R\'eville}
\affiliation{IRAP, Universit\'e Toulouse III - Paul Sabatier,
CNRS, CNES, Toulouse, France}

\author{Philippe Louarn}
\affiliation{IRAP, Universit\'e Toulouse III - Paul Sabatier,
CNRS, CNES, Toulouse, France}

\author{Yihong Wu}
\affiliation{IRAP, Universit\'e Toulouse III - Paul Sabatier, CNRS, CNES, Toulouse, France}

\author[0000-0001-8929-4006]{K\'{e}vin Dalmasse}
\affiliation{IRAP, Universit\'e Toulouse III - Paul Sabatier, CNRS, CNES, Toulouse, France}

\author{Vincent G\'{e}not}
\affiliation{IRAP, Universit\'e Toulouse III - Paul Sabatier,
CNRS, CNES, Toulouse, France}

\author[0000-0002-3520-4041]{Anthony W. Case}
\affiliation{Smithsonian Astrophysical Observatory, Cambridge, MA 02138 USA}

\author{Phyllis Whittlesey}
\affiliation{Space Sciences Laboratory, University of California, Berkeley, CA 94720-7450, USA}

\author{Davin Larson}
\affiliation{Space Sciences Laboratory, University of California, Berkeley, CA 94720-7450, USA}

\author[0000-0001-5258-6128]{Jasper S. Halekas}
\affiliation{Department of Physics and Astronomy, University of Iowa, IA 52242, USA}

\author{Roberto Livi}
\affiliation{University of California, Space Science Laboratory, Berkeley, CA, USA}

\author{Keith Goetz}
\affiliation{School of Physics and Astronomy, University of Minnesota, Minneapolis, MN 55455, USA}

\author{Peter R. Harvey}
\affiliation{Space Sciences Laboratory, University of California, Berkeley, CA 94720-7450, USA}

\author[0000-0003-3112-4201]{Robert J. MacDowall}
\affiliation{NASA Goddard Space Flight Center, Greenbelt, USA}

\author[0000-0003-1191-1558]{David Malaspina}
\affiliation{University of Colorado, Boulder, Laboratory for Atmospheric and Space Physics, Boulder, CO, USA}

\author[0000-0002-1573-7457]{Marc Pulupa}
\affiliation{Space Sciences Laboratory, University of California, Berkeley, CA 94720-7450, USA}

\author[0000-0002-0675-7907]{John Bonnell}
\affiliation{Space Sciences Laboratory, University of California, Berkeley, CA 94720-7450, USA}

\author[0000-0002-4401-0943]{Thierry Dudok de Witt}
\affil{LPC2E, CNRS and University of Orl\'eans, Orl\'eans, France}

\author{Emmanuel Penou}
\affiliation{IRAP, Universit\'e Toulouse III - Paul Sabatier, CNRS, CNES, Toulouse, France}



\begin{abstract}
The physical mechanisms that produce the slow solar wind are still highly debated. \textit{Parker Solar Probe's} (\textit{PSP}'s) second solar encounter provided a new opportunity to relate in situ measurements of the nascent slow solar wind with white-light images of streamer flows. We exploit data taken by the \textit{Solar and Heliospheric Observatory} (\textit{SOHO}), the \textit{Solar TErrestrial RElations Observatory} (\textit{STEREO}) and the Wide Imager on Solar Probe to reveal for the first time a close link between imaged streamer flows and the high-density plasma measured by the Solar Wind Electrons Alphas and Protons (SWEAP) experiment. We identify different types of slow winds measured by \textit{PSP} that we relate to the spacecraft's magnetic connectivity (or not) to streamer flows. SWEAP measured high-density and highly variable plasma when \textit{PSP} was well connected to streamers but more tenuous wind with much weaker density variations when it exited streamer flows. \textit{STEREO} imaging of the release and propagation of small transients from the Sun to \textit{PSP} reveals that the spacecraft was continually impacted by the southern edge of streamer transients. The impact of specific density structures is marked by a higher occurrence of magnetic field reversals measured by the FIELDS magnetometers. Magnetic reversals originating from the streamers are associated with larger density variations compared with reversals originating outside streamers. We tentatively interpret these findings in terms of magnetic reconnection between open magnetic fields and coronal loops with different properties, providing support for the formation of a subset of the slow wind by magnetic reconnection.
\end{abstract}



\section{Introduction} \label{sec:intro}

The solar wind plasma measured in situ has been classified into several different categories that could be related to different coronal sources \citep[e.g.][]{Xu2015JGRA}. The most clearly identified source regions visible in extreme ultraviolet (EUV) observations of the low corona are coronal holes. They host the footpoints of magnetic field lines connecting the corona to the interplanetary medium \citep[e.g][]{Krieger1973SoPh,Bame1976ApJ,Levine1977JGR}. There is currently no doubt that the fast and more tenuous solar wind originates in these coronal holes. \\

In contrast, the origin of the slow solar wind is less well understood. It could form by a number of distinct processes and from a number of different coronal structures including (1) the boundary of polar coronal holes and isolated low-latitude coronal holes \citep[e.g][]{Wang1994} and (2) small patches of open magnetic fields rooted in the vicinity of the magnetic loop complexes of active regions \citep[e.g][]{Kojima1999jgr,vanDriel2012SoPh,Culhane2014SoPh}. The slowest and densest solar wind measured in situ can be traced back to regions of the corona, called streamers, that appear very bright in white-light images  \citep{Sanchez2016JGRA,Sanchez2017ApJa}. Coronal rays are narrow lanes of enhanced brightness that extend from the corona to several tens of solar radii \citep{Druckmuller2014ApJ}. Coronal rays are referred to as ‘streamer rays’ when they originate in the vicinity of either helmet streamers or pseudo-streamers. Helmet streamers are systems of magnetic loops that separate open magnetic field lines of opposite magnetic polarity. The plasma escaping along these open magnetic field lines forms the helmet streamer rays. The polarity inversion line or coronal neutral line that forms near the tip of helmet streamers is the coronal origin of the heliospheric current sheet (HCS).  Helmet streamer rays are thought to engulf the HCS and be the source of the HPS typically measured in situ during crossings of the HCS \citep[][]{Winterhalter1994JGR}. Pseudo-streamers are coronal structures that separate open magnetic field lines of the same polarity; they produce streamer rays but do not produce a current sheet in the outer corona \citep[][]{Wang2007ApJ}. \\

The plasma outflows imaged along helmet streamer rays is highly intermittent and can be highly variable. A subset, at least, of these transient structures has been interpreted as outflowing magnetic flux ropes based on the analysis of multipoint imagery \citep{sheeley09} and the continuous tracking of these structures to their in situ counterparts \citep[][]{Rouillard2009SoPh, Rouillard2010ApJ, Rouillard2010JGRA, Rouillard2011ApJ}. Smaller scale density fluctuations detected in situ \citep{Viall2008jgr} in the slow wind have also been seen adjacent to small flux ropes \citep{kepko2016} and have been related to brightness variations in the corona \citep{Viall2015ApJ}. There is no doubt that a significant subset of the slow solar wind is composed of transient structures that form in the corona, many near the tip of helmet streamers where magnetic reconnection must occur, due to the presence of current sheets and null points \citep{Sanchez2017ApJa,Sanchez2017ApJb}. \\

The slow solar wind appears to originate from a very broad region of the corona, extending up to 40$^\circ$-50$^\circ$ away from the coronal neutral line. This suggests that the coronal conditions that produce the slow solar wind do not depend on the presence of a polarity inversion line. Background solar wind models that assume a coronal heating rate dependent on the local magnetic field properties lead to the interpretation that the slow solar wind is a natural consequence of the expansion rate of open magnetic field lines that channel the wind \citep[e.g.][]{Linker1999JGR,vanderHolst2010,Pinto2017ApJ}. The dependence of heating rates on the magnetic field properties find their justification in more detailed coronal heating models driven by Alfv\'en waves. None of these models are yet capable of simulating the composition of the slow solar wind, which would require a more dynamic mechanism involving reconnection between open and closed magnetic field lines \citep[e.g.][]{Baker2009}.\\

As already stated, the densest and slowest wind is traced back to the coronal neutral line where the bright streamer rays are formed. The thickness of these streamer rays can be measured in white-light images when the streamer is observed edge on. They typically extend over 10$^\circ$-20$^\circ$ in heliocentric latitude. Such a measurement represents a maximum thickness because even small latitudinal changes of the streamer belt would artificially broaden this region due to line-of-sight effects. This effect is analyzed using the Wide-field Imager for Solar PRobe (WISPR) data by \cite{Poirier2020ApJS} and also in the present paper. This angular extent is naturally much broader than that of the very thin and unresolved current sheet embedded in these coronal rays. The in situ counterparts of the coronal neutral line and the streamer rays are thought to be the HCS and the dense HPS. The HCS extends in situ over a heliocentric radial distance of just 1-10Mm while the HPS is about 500-700Mm \citep[][]{Winterhalter1994JGR}, which is on the upper end of the observed latitudinal extent of streamer rays, 10$^\circ$-20$^\circ$ when observed near 3\textit{R}$_\odot$. \\

Synthetic white-light images produced by three-dimensional (3D) coronal models provide a good representation of the extent of streamer rays and therefore the HPS \citep[e.g.][]{Pinto2017ApJ,Poirier2020ApJS}. In such simulations, the dense coronal regions result from the properties of magnetic field lines that are directly adjacent to the helmet streamer. Strong heating at the base of flux tubes, associated with strong footpoint field strengths, drives a high mass flux into the wind. In addition, the large expansion rate of flux tubes forces a rapid drop in the heating rate with altitude, preventing a strong acceleration of a dense wind \citep[e.g.][]{Wang1994,Pinto2017ApJ}. The excess density observed around the current sheet is related to a reconvergence of flux tubes near the top of helmet streamers. This produces a very slow and dense wind along the rays extending above streamer tops, which forms the HPS \citep[e.g.][]{Wang1994,Pinto2017ApJ}. \\

In addition to the intermittent eruption of helical magnetic fields already identified along streamer rays \citep[][]{Rouillard2009SoPh,Rouillard2010ApJ,Rouillard2010JGRA,Rouillard2011ApJ} and the continuous outflow of a very dense and slow background solar wind \citep[e.g.][]{Pinto2017ApJ}, we expect other dynamic processes to perturb the solar wind from streamers. These include magnetic reconnection between coronal loops and open magnetic fields that connect to the streamer tops or, instead, between open magnetic field lines of opposite polarities that meet at the streamer tops. This should produce transient outflows with distinct magnetic signatures and shears that would modify the properties of the wind expelled from streamer rays \citep[e.g.][]{Owens2018ApJ}. \\

Due to the large distances between the regions imaged by coronagraphs and in situ measurements taken mostly near 1 au (astronomical unit), streamer flows that fade by 60\,--\,70 solar radii could not be related clearly with in situ measurements until now. Previous analyses based on heliospheric imaging were limited to tracking only those streamer transients from the corona to 1 au that become swept up by high-speed streams. The advent of \textit{PSP} and its measurements of the background solar wind as close as 35 solar radii \citep[][]{Bale2019Nat,Kasper2019Nat} alleviate some of these past difficulties. Even white-light features that disappear by 50\,--\,70 solar radii in the heliospheric images \citep{Eyles2009SoPh} taken by \textit{STEREO} can be detected by the \textit{PSP} plasma and magnetic field detectors before they are no longer discernible in the images. In the present paper, we exploit multipoint data during \textit{PSP}'s second solar encounter. At this time, \textit{STEREO-A} (\textit{STA}) was optimally located to track density variations continuously from the Sun to \textit{PSP}.\\

The present work is structured as follows. We start by presenting the observational context from both a remote-sensing and in situ perspective. We then present \textit{STEREO} and \textit{PSP} images of bright structures expelled by helmet streamers in the direction of \textit{PSP}. In the third part, we compare these images with the in situ measurements from \textit{PSP} at the predicted times of impact of the imaged density structures and reveal the presence of multiple switchbacks. Finally, we discuss the possible origins of these features at the Sun by considering \textit{STEREO} EUV images.

\section{Orbital details of the second \textit{PSP} encounter} \label{sec:OrbPSP}

\begin{figure*}[ht!]
\centering
\includegraphics[scale=0.32]{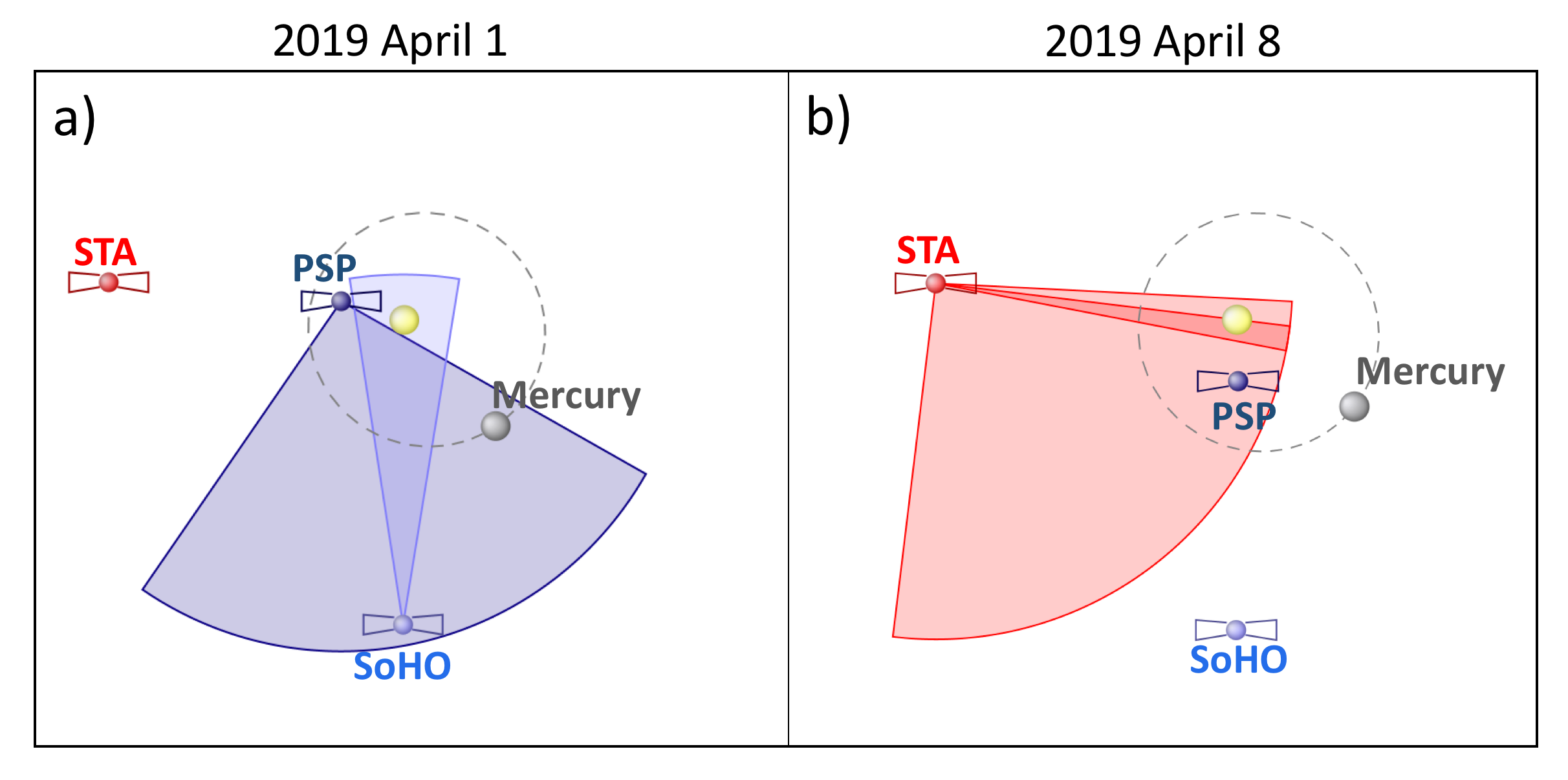}
\caption{Two views of the ecliptic plane from solar north showing the orbital positions of \textit{STA}, \textit{PSP} and \textit{SOHO} in inertial coordinates (Heliocentric Aries Ecliptic) on 2019 April 1 and 8. The fields of view of the combined \textit{PSP} WISPR, \textit{SOHO} LASCO C3 and \textit{STA} COR2/HI1 instruments used in this study are shown as dark blue, light blue, and red shaded areas, respectively. The position of Mercury is also shown as a gray disk. This figure was produced using the Propagation Tool described in \citet{Rouillard2017}. \label{fig:3DPos&FOV}}
\end{figure*}

The second \textit{PSP} solar encounter occurred between 2019 March 30 and 2019 April 10. Figure~\ref{fig:3DPos&FOV} presents views of the ecliptic plane from solar north, inside this time window, on April 1 and 8. The figure also shows the combined fields of view of the WISPR \citep[][]{Vourlidas2016SSRv} instruments on \textit{PSP} (shown as a shaded blue area), the \textit{SOHO} LASCO C3 instrument \citep{Brueckner1995} as well as the Outer CORonagraph (COR2) and the combined Heliospheric Imagers (HI1 and HI2) instruments (red shaded areas) on board \textit{STA}. The latter instruments form part of the Sun-Earth Connection Coronal and Heliospheric Investigation (SECCHI) package \citep[][]{Howard2008SSRv}. \\

\textit{PSP} is located off the east limb of the Sun just outside the outer edge of the \textit{SOHO} C3 field of view. On April 8, \textit{PSP} was situated inside the field of view of the \textit{STA} HIs that were imaging plasma off the west limb of the Sun viewed from \textit{STA}. Plasma that escaped the Sun could have been imaged by SECCHI before it was measured \textit{in situ} by \textit{PSP}.  \\

\begin{figure}[ht!]
\includegraphics[scale=0.58]{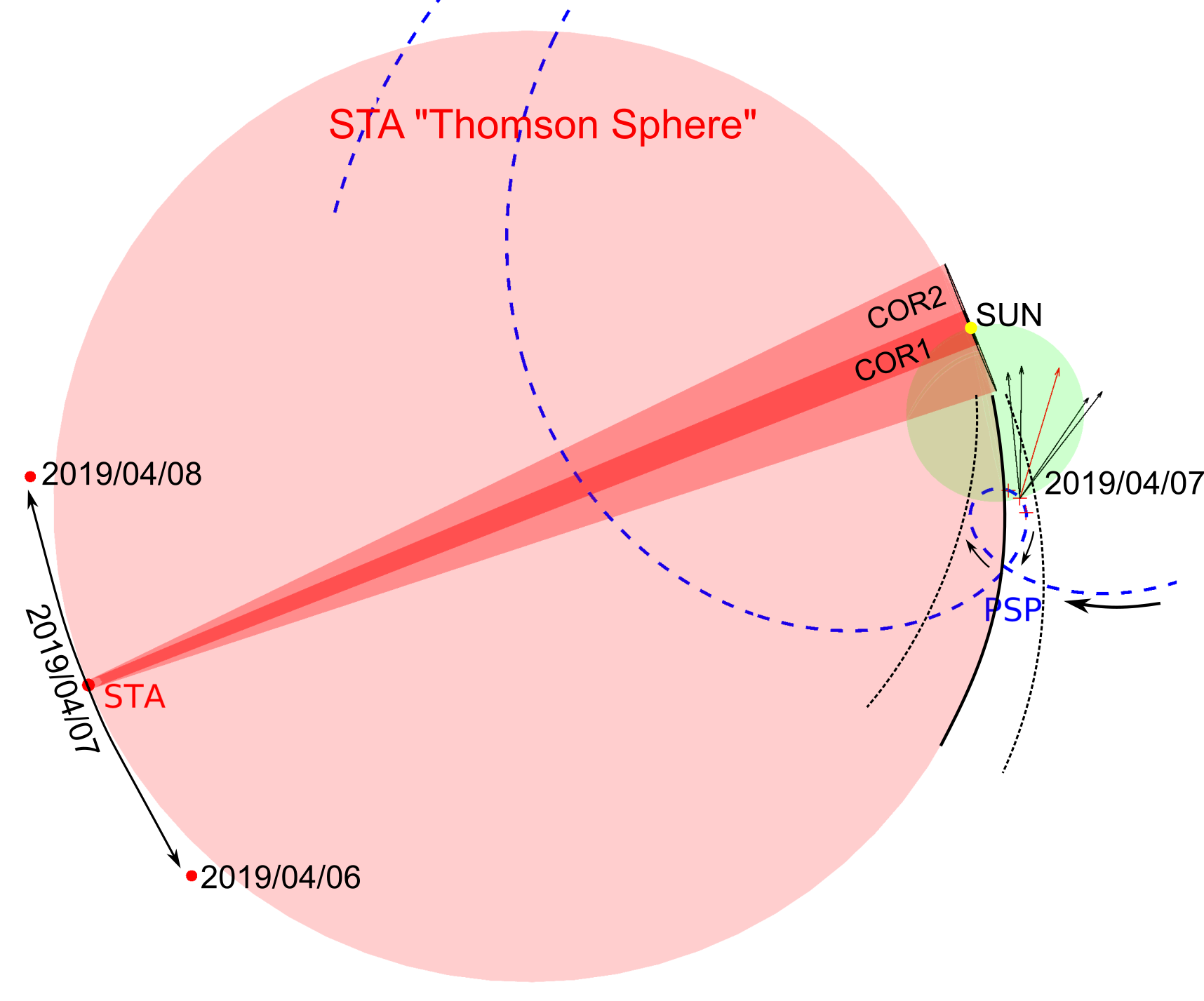}
\caption{A view of the ecliptic plane from solar north showing the relative positions of the Sun,  \textit{STA} (red), and \textit{PSP} (blue) in Carrington coordinates on 2019 April 7. In this coordinate system, the orbit of \textit{PSP}, shown as a dashed blue line, executes a loop near perihelion. The Thomson sphere of \textit{STA} is shown as the red disk and that of \textit{PSP} as a green disk. The fields of view of \textit{STA} COR1 and COR2 depicted with different shades of red and the area swept by the extent of the field of view of HI1 can be seen as the black curve on \textit{STA}'s Thomson sphere. This curve sweeps an area bound by two similar black dotted curves between April 6 and April 8, the time interval when ejections are analyzed in this paper. These black dotted curves bound the entire perihelion passage, meaning that density structures can be tracked near the Thomson sphere of \textit{STA} continuously from the Sun to \textit{PSP} at these times.  \label{fig:PossCARR}}
\end{figure} 

\begin{figure*}[ht!]
\centering
\includegraphics[scale=0.8]{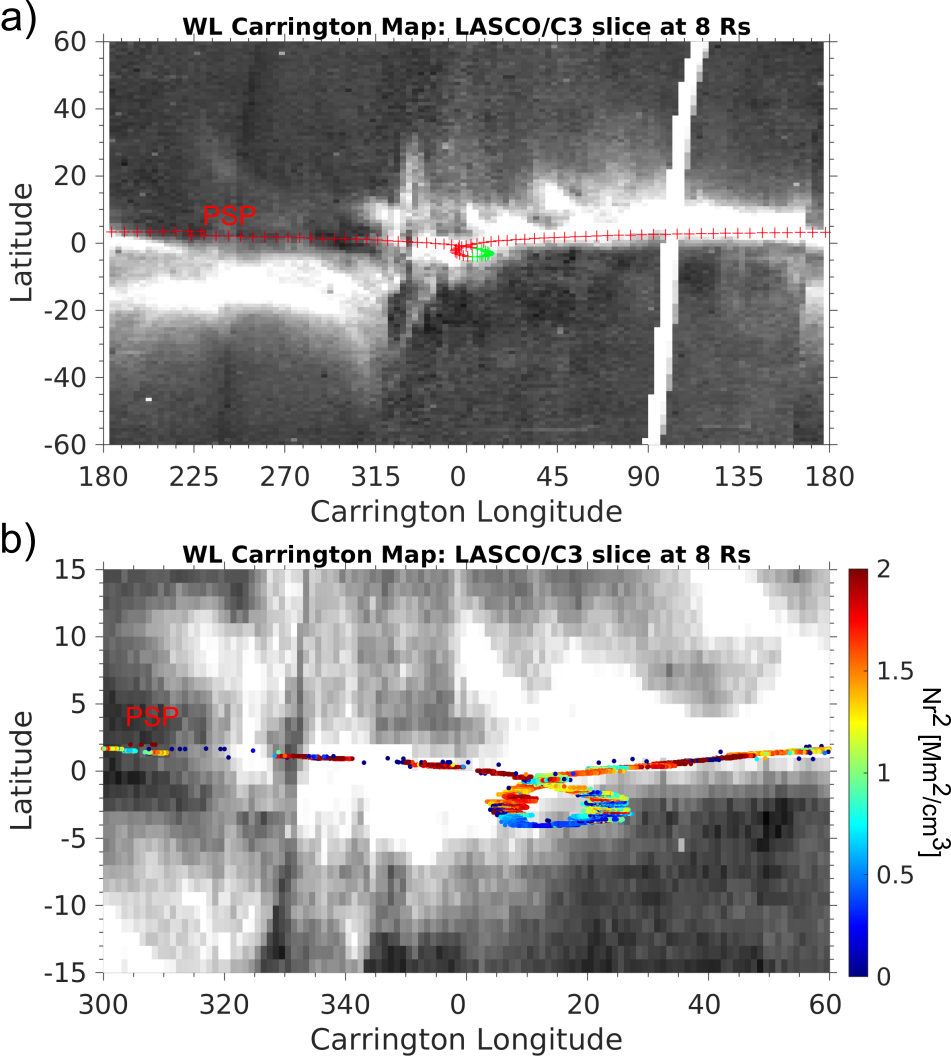}
\caption{A comparison of the relative heliographic positions of \textit{PSP} and streamer rays at 8\textit{R}$_\odot$. Panel (a): a full Carrington map constructed by extracting pixels of LASCO C3 images along the arc of a circle centered at the Sun center and passing at 8\textit{R}$_\odot$ off the east solar limb as viewed from \textit{SOHO}. The heliographic coordinates of \textit{PSP} (without accounting for propagation effects) is shown as the red crosses. The part of the orbit that we focus on here is highlighted with green crosses. Panel (b): a zoomed-in view of the same Carrington map but focusing on the encounter period. The \textit{PSP} path corresponds to the points of magnetic connectivity traced back to the radial distance of the map (8\textit{R}$_\odot$). The connectivity is estimated by assuming the magnetic field follows a Parker spiral calculated from the speed of the solar wind measured in situ at \textit{PSP}. The color coding is defined by the density ($N r^2$) measured \textit{in situ} by \textit{PSP} with red corresponding to high densities and blue to low densities. 
 \label{fig:carrmap}}
\end{figure*}

We can illustrate this observational capability in more detail by changing from an inertial to a Sun-corotating frame such as Carrington coordinates. This representation is shown in Figure~\ref{fig:PossCARR} for the time interval of interest here. The theory of Thomson scattering tells us that coronal regions located near the ``Thomson sphere'' contribute most to the visible light recorded by an imager \citep{Vourlidas2006ApJ}. Figure~\ref{fig:PossCARR} illustrates the total ecliptic area observed by the section of the Thomson sphere inside the field of view of the \textit{STA} (COR2/HI1) and WISPR-I instruments of \textit{PSP} and \textit{STA} from 2019 April 5 to 10. The WISPR instruments consist of two cameras, in the inner (WISPR-I) and outer (WISPR-O) imagers; in this study, we make use of WISPR-I. WISPR-I extends in elongation angles from 13.5$^\circ$ to 53$^\circ$ and WISPR-O extends from 50$^\circ$ to 108$^\circ$. \\

Figure~\ref{fig:PossCARR} also shows that \textit{PSP}'s orbit remained near the Thomson sphere of \textit{STA} for an extended period from April 6 to April 9. The Carrington longitude of \textit{PSP} only changed by 4.5$^\circ$ between April 5 and April 11, moving from 7.6$^\circ$ to 12.1$^\circ$ longitude, respectively. \textit{PSP} was making in situ measurements in almost the same region during the period we focus on here.


\section{Relating streamer rays to the dense solar wind} \label{sec:raysinsitu}

To provide further context to the analysis that follows, we present in Figure~\ref{fig:carrmap}a a Carrington map obtained from LASCO C3 observations on board \textit{SOHO} during a whole solar rotation. This Carrington map is constructed by taking a band of pixels at a given heliocentric radial distance from each LASCO C3 image to produce a latitudinal strip. Each strip is then assigned a Carrington longitude by assuming that the observed brightness comes from the Thomson sphere of the instrument. This representation provides a powerful way of visualizing the global structure of the streamers. The bright band observed at low latitude and near the equator, and extending over all longitudes, is the streamer belt that corresponds to the densest regions of the corona. As discussed in the introduction, the latitudinal width of the band is about 10$^\circ$\,--\,20$^\circ$. \\

The trajectory of \textit{PSP} is overplotted as red/green crosses on this map. The periods of corotation and superrotation can be seen as the small loop near Carrington longitudes 355$^\circ$\,--\,10$^\circ$ (near the center of the map). The map reveals that \textit{PSP} remained near the edge of the streamer belt throughout its second encounter. As \textit{PSP} did not cross the center of the streamer until well after perihelion, we expect from Figure~\ref{fig:carrmap} that the probe remained in the same magnetic sector for most of the second encounter. In situ measurements of the magnetic field confirm that \textit{PSP} only crossed the polarity inversion line on around April 16.\\



\begin{figure*}[ht!]
\centering
\includegraphics[scale=0.7]{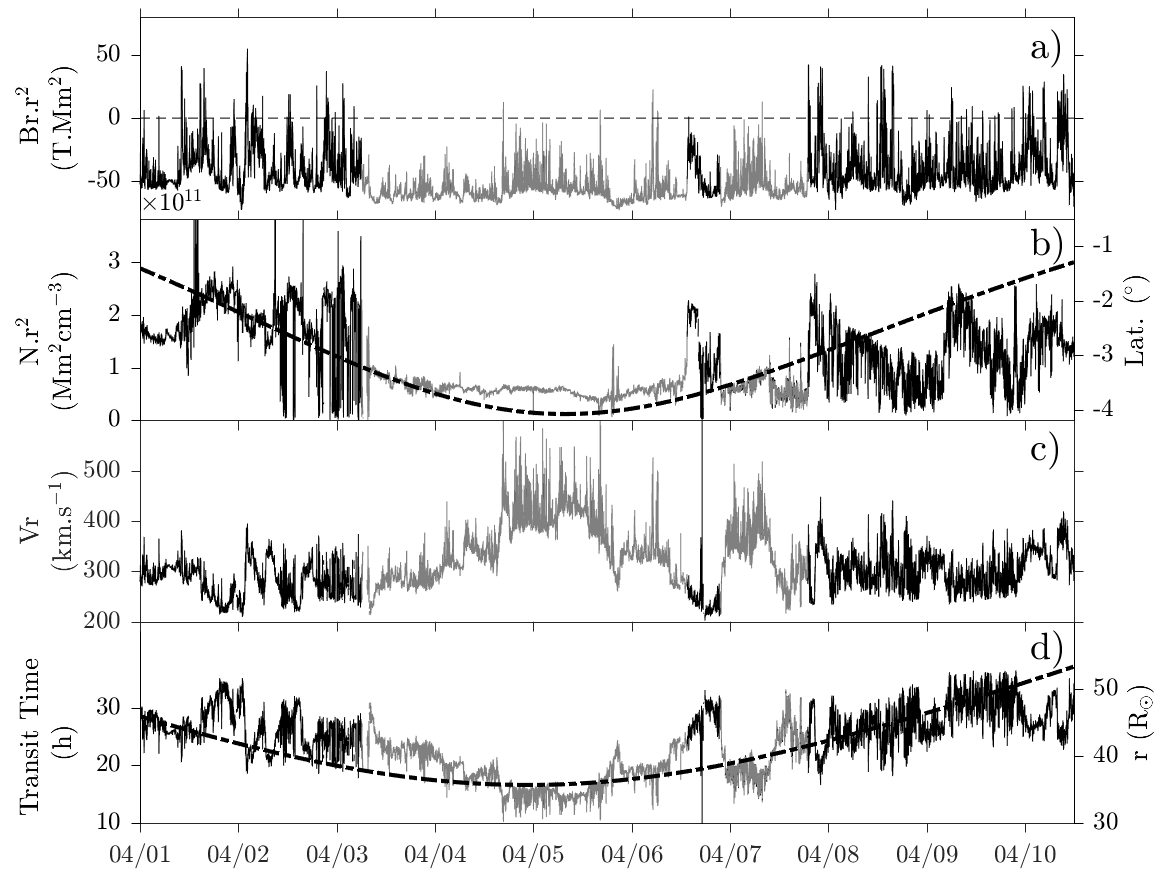}
\caption{In situ measurements taken by the FIELDS and SWEAP instrument suites on \textit{PSP} over a 10 day interval centered around \textit{PSP}'s second perihelion. Panel (a): the measured radial component of the magnetic field ($B_r$) in radial tangential normal (RTN) coordinates, multiplied by the square of the radial distance ($r^2$). Panel (b): proton density of the solar wind ($N$) also multiplied by $r^2$. The heliographic latitude (Lat ($^\circ$)) of \textit{PSP} is also plotted as a dash-dotted line. Panel (c): the radial speed of solar wind protons ($V_r$). Panel (d): the plasma transit time from the 2.5\textit{R}$_\odot$ to \textit{PSP} in hours (hr) computed from $V_r$. The heliocentric radial distance of \textit{PSP} is also shown as a dash-dotted line. The plasma measurements associated with streamer flows in this study are shown in black while those from outside streamers are shown in gray. The in situ data were averaged over 30 seconds.
 \label{fig:contextinsitu}}
\end{figure*}

We present in Figure~\ref{fig:contextinsitu} a summary plot of in situ measurements made by \textit{PSP} over a 10 day period centered on perihelion. The FIELDS suite of instruments provided combined measurements of magnetic and electric fields \citep[][]{Bale2016SSRv}. Magnetic fields are measured using both fluxgate and search-coil (induction) magnetometers mounted on a deployable boom in the spacecraft umbra. We here show measurements made by the magnetometer.  As highlighted in \cite{Bale2019Nat}, the FIELDS magnetometer measured magnetic fields with predominantly Sun-pointing polarity throughout the second encounter. This is reflected in panel (a), where the radial field component remains negative. As discussed by \cite{Bale2019Nat} and \cite{Kasper2019Nat}, the magnetic field exhibited sudden short-lived reversals of the magnetic field direction that were not associated with correlated changes in the pitch angle of suprathermal electrons. Hence, the polarity of magnetic fields does not change in these structures, and the reversals are interpreted as ``folds'' or ``switchbacks'' in the magnetic field lines \citep{Bale2019Nat}. These are clearly seen in panel (a). We plot here the radial magnetic field multiplied by the square of the heliocentric radial distance of the spacecraft ($\mathrm{B_r r^2}$) to remove the effect of the varying heliocentric distance of the spacecraft. \\

The density and speed of the solar wind protons are measured by the Solar Probe Cup (SPC) part of the instrument suite of the Solar Wind Electrons Alphas and Protons \citep[SWEAP; ][]{Kasper2016SSRv} experiment. The Solar Probe Cup has a 60$^\circ$  Sun-pointing (full-width) field of view and is placed near the edge of the probe's heat shield. The operating principle of the instrument is described in \citet{Case_2019} and is similar to that of previous Faraday Cup experiments in space. The instrument measures the current deposited by inflowing ions (or electrons) onto a segmented metal plate at the base of the cup, and those charge carriers are discriminated with respect to their kinetic energy per charge using a set of transparent, high-voltage grids to which an A/C waveform is applied. The A/C waveform is stepped through a series of subranges in voltage spanning the energy of the bulk solar wind, and the corresponding amplitudes of the modulated current are used to reconstruct the radial kinetic energy-per-charge distribution function for the solar wind. The proton density, temperature, and velocity moments are derived from direct integration of the measured distribution in the neighborhood of the primary peak in the ion current \citep{Case_2019}. The 30 second averages of proton densities and radial speeds derived from SPC data are shown in panels (b) and (c) of Figure \ref{fig:contextinsitu}. The proton densities have been multiplied by $\mathrm{r^2}$ in panel (b). Like the magnetic field data, the proton densities also display considerable variability. The SPC appears to have measured different regimes of solar wind during the encounter, with periods of dense, highly variable, and very slow solar wind ($\leq$300 km/s) and less dense and faster plasma ($>$300 km/s) close to perihelion. We have used different shades of black on this panel to highlight these different regimes. We also overplot in panel (b) the rapidly changing heliographic latitude of the spacecraft. We interpret these three regimes as a consequence of the changing latitude of \textit{PSP}, which temporarily exits streamer flows. \\

To test this idea, Figure~\ref{fig:carrmap}b provides a zoomed-in view of panel (a) around perihelion. In Figure~\ref{fig:carrmap}b, we trace back the magnetic connectivity of \textit{PSP} to the height at which the map is constructed, e.g. 8\textit{R}$_\odot$, by using the solar wind speed measured \textit{in situ} by \textit{PSP} (see Figure ~\ref{fig:contextinsitu}c). \textit{PSP}'s movement of a few degrees in latitude is clearly seen in this zoomed-in view (Figure~\ref{fig:carrmap}b). The map reveals that \textit{PSP} temporarily moved away from the center of the streamer into a less bright region before rapidly re-entering the streamer. To compare directly the brightness in our map with the measured plasma densities, we have color-coded the trajectory of \textit{PSP} according to the density ($Nr^2$) measured \textit{in situ}. We can clearly see that the elevated density measured by \textit{PSP} corresponds to periods when the spacecraft is connected to the streamer flows. The lowest plasma density corresponds to periods when \textit{PSP} exits the streamer temporarily. Such a close match between plasma flows measured \textit{in situ} and coronal images is unprecedented and is, of course, related to the proximity of \textit{PSP} to the corona, with the spacecraft only about 20\textit{R}$_\odot$ away from the imaged streamer rays. \\

\begin{figure}[ht!]
\centering
\includegraphics[scale=0.45]{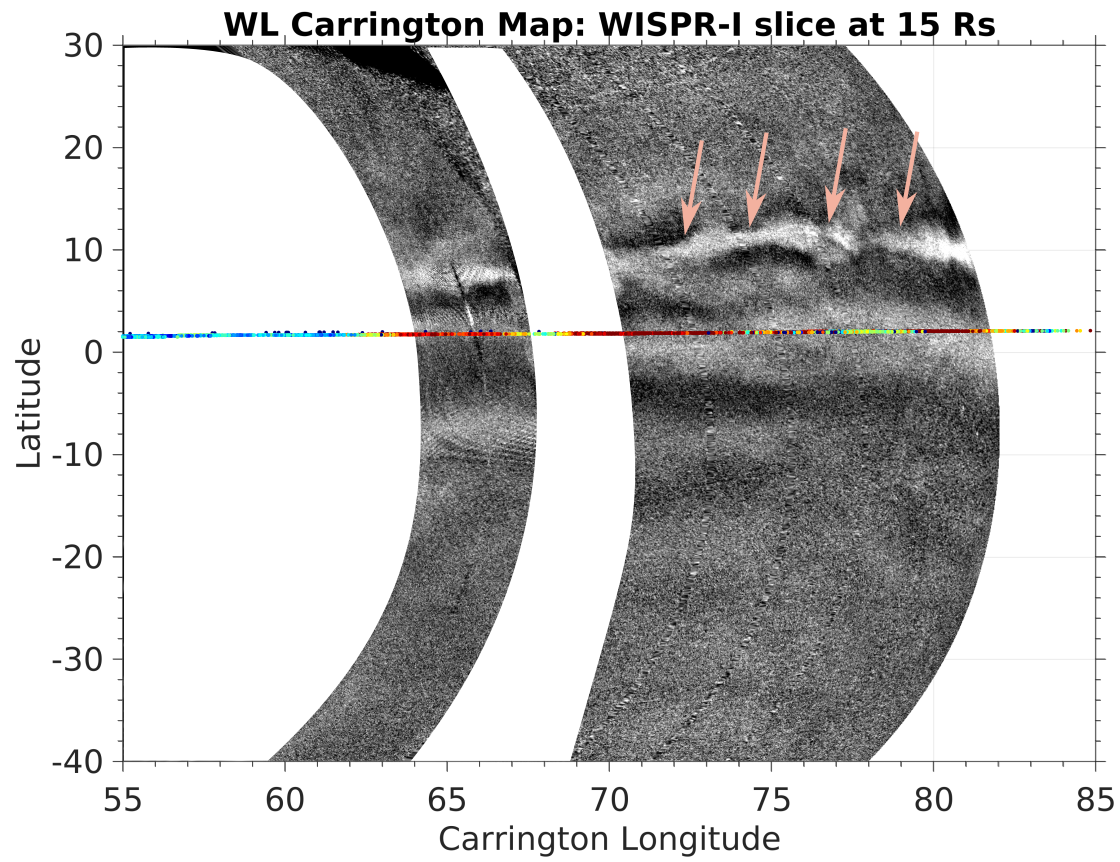}
\caption{Similar to Figure \ref{fig:carrmap} but for WISPR-I observations and limited to the range of Carrington longitudes imaged by WISPR-I (between 55$^\circ$ and 85$^\circ$). The colored line indicates the heliographic coordinates at the radial distance of the map (15\textit{R}$_\odot$) of the Parker spiral connected to \textit{PSP}. Its color coding is defined by the normalized density ($N r^2$) measured \textit{in situ} by \textit{PSP} with red corresponding to high densities and blue to low densities. The colored arrows mark the likely location of the coronal neutral line. 
 \label{fig:carrmapW}}
\end{figure}

Figure \ref{fig:carrmapW} presents a similar map but constructed using WISPR-I images taken between April 1 and 8. WISPR-I is a white-light instrument that observes and records the brightness of the F-corona, produced by light scattered off dust \citep{stenborg_2018}, and the K-corona produced from the scattering of photospheric light by coronal and solar wind electrons. The Level-1 FITS files contain brightness measurements that must be normalized for exposure time and then corrected for the vignetting effects of the detector. The vignetting function and calibration constant were both initially determined during preflight calibration of the instrument but have since been modified based on stellar photometry using techniques adapted from \cite{Bewsher2012} and \cite{tappin_2015}. The signal of the F-corona was removed by adapting a technique developed by \citet{stenborg_2018} on SECCHI HI1 data. This technique was applied to all the WISPR-I data to produce images of the K-corona. \\

The technique used to build the Carrington from WISPR-I images is described in detail in \cite{Poirier2020ApJS}. A search is made in each image of the K-corona for all pixels associated with lines of sight intersecting the Thomson sphere at a heliocentric radial distance of 15\textit{R}$_\odot$. As highlighted by Figure~\ref{fig:PossCARR}, WISPR-I provides a zoomed-in view of the corona. The WISPR-I data in Figure~\ref{fig:carrmapW} cover a small range of Carrington longitudes from 55$^\circ$ to 85$^\circ$. The bright band of pixels that we associated with streamer rays in the LASCO C3 Carrington map extended from $\sim-5^\circ$ to about 10$^\circ$ between Carrington longitudes 40$^\circ$ and 80$^\circ$; this matches the latitudinal band where we see the brightest rays in the WISPR-I map in Figure~\ref{fig:carrmapW}. Similar to the maps produced for the first encounter and analyzed in detail by \cite{Poirier2020ApJS}, the WISPR-I images reveal substructure in the streamer rays that is not visible in LASCO (Figure~\ref{fig:carrmap}) or \textit{STEREO} images. \\

The plasma imaged by WISPR-I over the range of longitudes shown in Figure \ref{fig:carrmapW} is related to source regions that released plasma toward \textit{PSP} between March 23 and 27, well before the start of the second encounter. Hence, we cannot make the one-to-one association between WISPR-I imagery and \textit{PSP} in situ data; this is possible with \textit{STA} images, as shown later in this paper.\\

We plot on this map the normalized density ($Nr^2$) measured in situ by \textit{PSP} with the same color scheme as in Figure \ref{fig:carrmap}. The plasma densities are very elevated during this time interval because the probe is passing near the southern edge of the streamer flows. The brightest rays marked by arrows in this map correspond to the densest part of the streamer, where the current sheet is likely to be located. If this is the case, then Figure \ref{fig:carrmapW} reveals that \textit{PSP} passed just 4$^\circ$-10$^\circ$ south of the HCS during this time. The in situ measurements confirm that \textit{PSP} remained in the negative polarity during this time interval and did not cross the HCS \citep{Bale2019Nat}. The flows are perturbed by bursts of switchbacks (Figure \ref{fig:contextinsitu}). \\

We conclude from these preliminary studies that \textit{PSP} primarily sampled plasma escaping from the southern flank of the streamer; therefore, the probe remained in a single magnetic sector (inward-pointing magnetic fields) throughout the second encounter and, in particular, during the period of interest here. The plasma density normalized by $r^2$  inside the streamer flows is up to a factor of 6 higher than that in the slow wind emerging outside streamers. Near perihelion, \textit{PSP} temporarily exited the streamer to sample more tenuous slow solar wind emerging from another source that could be associated with a region just inside the outer edge of a coronal hole. We find evidence that the reversals of the magnetic field lines detected by FIELDS occur in bursts or clumps when they originate in streamer flows, and are sometimes accompanied by significant density variations. The densest plasma measured by SPC exhibits great variability, with changes of normalized density of a factor of 3; such fluctuations would easily be detected as strong brightness variations by coronagraphs and heliospheric imagers. In contrast, the switchbacks that occur in the slightly faster and more tenuous solar wind outside streamer flows are shorter lived and do not exhibit strong density variations. These should remain undetected of current white-light instruments.\\

We now investigate the origin of the density structures detected by \textit{PSP} between April 6 and 11 as it re-entered streamer flows, by using multipoint coronal imaging. This period is of great interest because, as already stated, the plasma directed toward \textit{PSP} should have been imaged by \textit{STA}.

\section{Streamer activity captured by \textit{STEREO-A}} 
\label{sec:ImagingStreamers}

\begin{figure*}[ht!]
\centering
\includegraphics[scale=0.9]{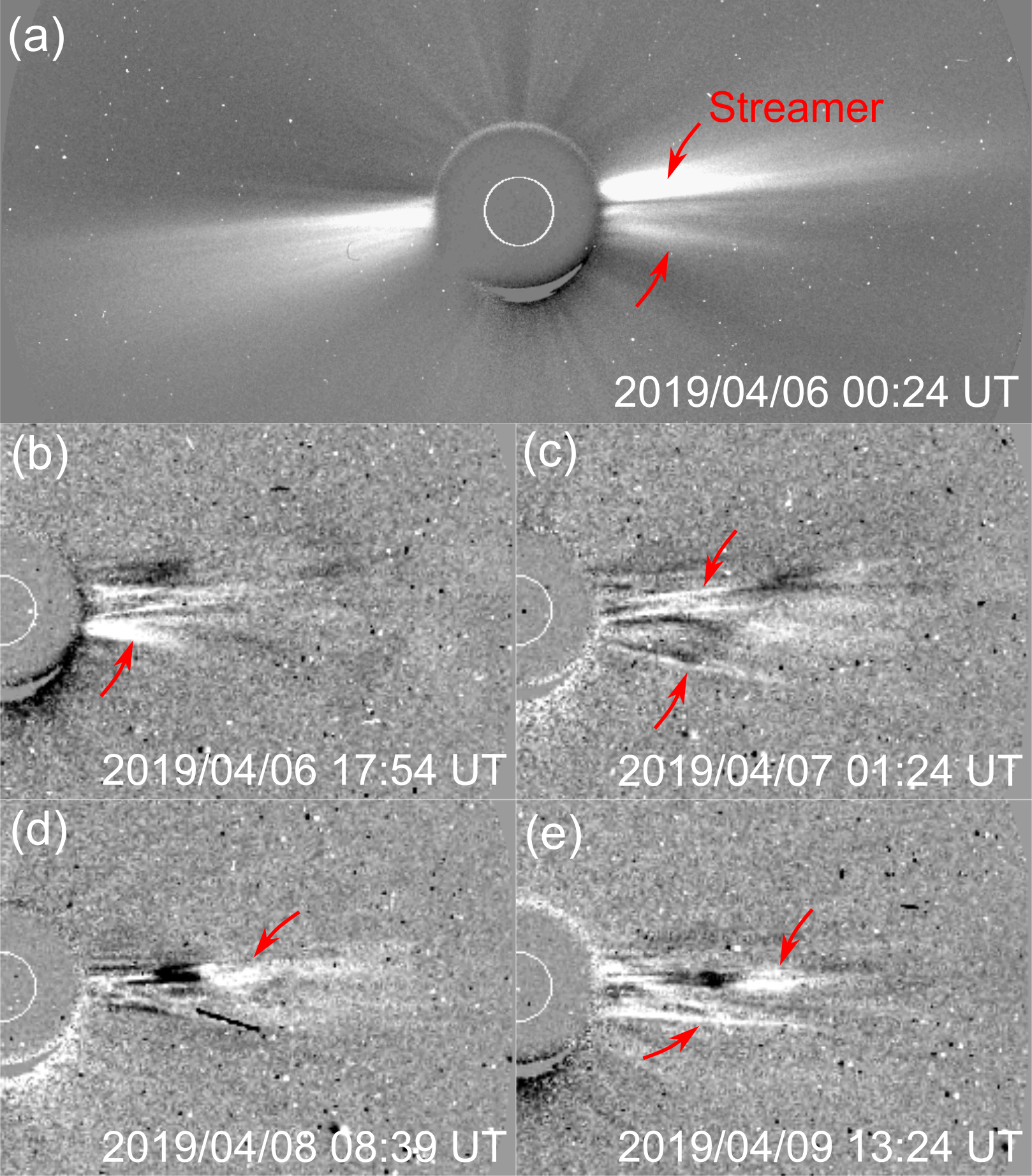}
\caption{Panel (a): a background-subtracted COR2A image from April 6, just after \textit{PSP}'s perihelion. We label the position of the bright west-limb streamer in red and mark additional streamer rays that suddenly appear in the image. Panels (b)-(e): running-difference images from COR2A throughout the time interval extending from April 6 to 9, during which we track density structures to \textit{PSP}. The colored arrows mark the positions of the small transients in the images. \label{fig:STEREOCOR2_Sample}}
\end{figure*}

We begin this section by examining coronal activity off the west limb of the Sun imaged by \textit{STA} COR2 and HI1. Figure \ref{fig:STEREOCOR2_Sample}a shows a COR2 image from 2019 April 6 00:24~UT where we have subtracted the background F-corona to reveal the K-corona. This image shows the presence of a streamer located a few degrees north of the equatorial plane and bright rays that start to appear just a few degrees south of that streamer. At these times, the position angle (PA) of \textit{PSP} lies along the southern rays,  south of the main streamer rays in this image. This is expected from examination of Figure \ref{fig:carrmap}b because, just after perihelion (right-hand side of the loop at 20$^\circ$ Carrington longitude), \textit{PSP} is located just south of the bright streamer and west of the portion of a streamer that is entering the plane of the sky of \textit{STA}.\\

In panels (b)-(e) of Figure~\ref{fig:STEREOCOR2_Sample}, we present four COR2A running-difference images from the period April 6 to 9, at times when small-scale transient structures were ejected over the west solar limb. Red arrows in those panels mark those ejecting dense structures (so-called ``blobs''). Some of the structures take the shapes of loops; others have V-shaped aspects. They are reminiscent of the ejection of helical magnetic fields \citep{sheeley09}. \\

\begin{figure}[h!]
\centering
\includegraphics[scale=1.05]{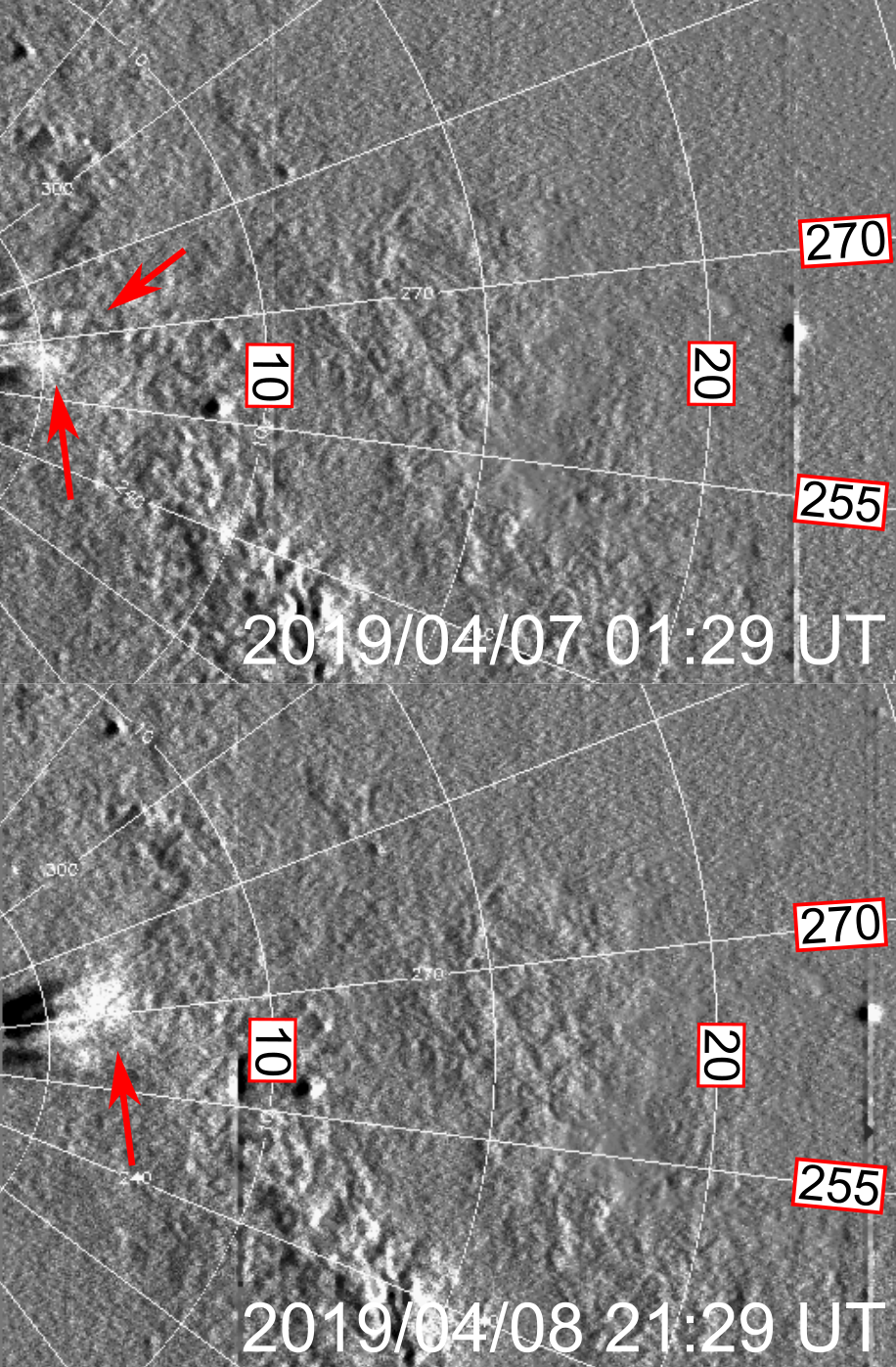}
\caption{Two running-difference images from HI1A showing the bright features tracked to \textit{PSP}. \textit{PSP} is situated at a PA of 265$^\circ$ and near the 10$^\circ$ elongation mark at the time. The colored arrows mark the positions of the small transients in the images. \label{fig:STEREOHI1_Sample}}
\end{figure}

Most of the streamer outflows could be traced into HI1 images, and a subset as far out as HI2. Figure~\ref{fig:STEREOHI1_Sample} presents two HI1 running-difference images that show some of the streamer outflows. Based on the \textit{PSP} orbit shown in Figure~\ref{fig:PossCARR}, HI1 imagery offers a unique opportunity to track bright features continuously out to \textit{PSP}; \textit{PSP} was located at PA=265$^\circ$ (i.e. between the two PAs labeled in the Figure~\ref{fig:STEREOHI1_Sample}). There is a constant ejection of blobs near the equatorial plane throughout the interval of interest. Most of the outflows seem to have a width lower than 15$^\circ$ and a speed of around 350~$\mathrm{kms^{-1}}$. The ejection that extends most in PA (30$^\circ$) during that time interval was imaged by HI1 around 21:29~UT on April 8 (see Figure~\ref{fig:STEREOHI1_Sample} bottom panel). This blob exhibits V-shape structures and could consist of helical magnetic fields. The central axis feature propagates along PA=270$^\circ$ and, as we shall see, \textit{PSP} could have been impacted by its southern edge.  \\

\section{Tracing density structures from \textit{PSP} back to the Sun} \label{sec:Jmaps}

The representation of white-light imagery in the form of time-elongation (or time-height) maps provides a powerful way to track the evolution of coronal structures moving through the optically thin solar atmosphere. These maps, traditionally called J‐maps, were first produced with LASCO C2/C3 images \citep{sheeley99} and were subsequently adapted to \textit{STEREO} COR and HI images \citep{sheeley08, Davies2009}. J-maps are constructed by extracting strips along a fixed PA from a sequence of coronagraph and/or heliospheric images. The extracted strips are plotted vertically as a function of time to generate an elongation versus time map. J-maps based on observations from near 1 au are typically built from running-difference images to minimize the contribution of the F-corona and to highlight faint propagating features. 

\begin{figure*}[ht!]
\centering
\includegraphics[scale=0.333]{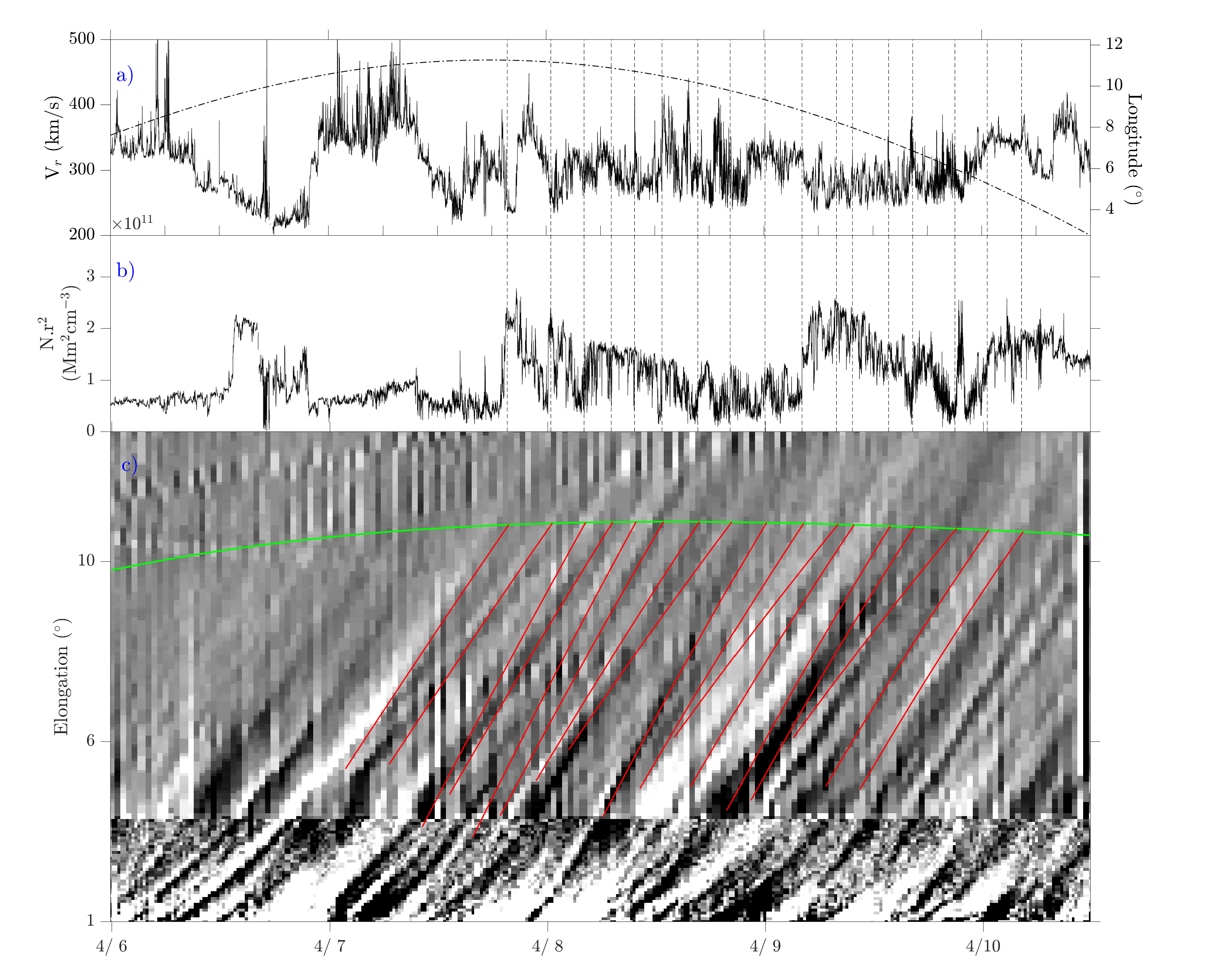}
\caption{Panels (a) and (b): the radial component of the solar wind speed and the density measured by SPC. The Carrington longitude of \textit{PSP} is shown as a dash-dotted line in panel (a). The vertical dashed lines plotted in both panels (a) and (b) mark the onset of noticeable density variations. Panel (c): a J-map constructed from COR2A and HI1A images. The green curve shows the elongation variation of \textit{PSP} during this time interval. The red curves are the time-elongation profiles of density structures marked by the vertical bars in panel (b), assuming each of the structure propagates from the Sun to \textit{PSP} at the speed measured in situ (panel (a)).  \label{fig:JmapSTEREO}}
\end{figure*}

Figure~\ref{fig:JmapSTEREO} presents such a J-map, derived from COR2A and HI1A images taken between 2019 April 6 and 11, along $\mathrm{PA=265^\circ}$). This PA was chosen to track features directed toward \textit{PSP}'s position. Because white-light images are integrated along the line of sight, features directed along \textit{PSP}'s PA are not necessarily directed toward \textit{PSP}. COR2A observations extend out to an elongation of approximately 4$^\circ$, while HI1 observations extend $\sim$4$^\circ$-24$^\circ$. From April 6 to 11, \textit{PSP} was at an elongation (as viewed from \textit{STA}) that varied from 9.9$^\circ$ to 11$^\circ$. The J-map shown in Figure~\ref{fig:JmapSTEREO} covers the entire field of view of COR2A and about half of HI1A's field of view, extending out just beyond the maximum elongation reached by \textit{PSP} during the interval of interest (11$^\circ$). 

The J-map confirms that, during this time interval, a profusion of density structures erupted from the corona along the PA of \textit{PSP}. This activity is particularly intense after April 7 (tick label 4/7). As can be seen in Figures \ref{fig:STEREOCOR2_Sample} and \ref{fig:STEREOHI1_Sample}, a significant burst of small eruptions occurs between 12UT on April 7 and 00UT on April 8. The shape of the tracks in the J-map shows that the density structures accelerate in the coronagraph field of view and then maintain a more constant progression through the inner part of the HI1A field of view, all the way to \textit{PSP}.  \\

In order to connect \textit{PSP} measurements of sudden density changes with the inclined tracks seen in the J-maps, we plot in panels (a) and (b) of Figure~\ref{fig:JmapSTEREO}, respectively, the radial speed and density of the solar wind plasma measured in situ by \textit{PSP}. We mark some of the most notable density variations measured by \textit{PSP} during that time interval with vertical dashed lines. For each impact, we know the measured radial speed of the density structure ($V_r$), the heliocentric radial distance of \textit{PSP} ($r_p$), and the longitudinal separation between \textit{STA} and \textit{PSP} ($\beta$). The latter changed from 68$^\circ$ to 124$^\circ$ during the interval spanned by the J-map. We then used the approach of \cite{Rouillard2010ApJ} to compute the apparent elongation variation ($\alpha(t)$) that each density feature would show in the J-map if moving radially outwards at a constant speed of $V_r$:

\begin{equation}
    tan[\alpha(t)]=\frac{V_rt sin(\beta)}{r_A-V_r t cos(\beta)}
    \label{eq:fixedphi}
\end{equation}

where time $t$ runs backwards from the time of impact at \textit{PSP} and $r_A$ is the radial distance of \textit{STA}. We assume that $\beta$ remains constant during the propagation of each structure, in other words that \textit{STA}'s longitude remains constant and that the radially outflowing features do not corotate with the Sun but, instead, moved in a purely heliocentric radial direction. SPC has detected a significant tangential component of the plasma velocity at its two first perihelia \citep{Kasper2019Nat}. Any effect of plasma motion on the shape of the determined tracks is left to be investigated in a future study.  \\ 

Nearly all of the density structures detected by \textit{PSP} in situ have an apparent track that matches an observed track in the \textit{STA} J-map. The varying inclination of the reconstructed tracks in the J-map is related to the corotation of the plasma source at the Sun. The density structures are expelled closer to the observer at the start of the interval than toward the end of the time interval.  By visual inspection, we find that the best match between the observed and traced tracks occurs after April 7. Before this time, the connection of the in situ and J-maps tracks is more ambiguous; this is likely related to the positions of \textit{PSP} with respect to the \textit{STA} plane of the sky (and also the Thomson sphere).   \\

We are able to relate the times of large density peaks measured by \textit{PSP} on April 7 19:30 UT and April 9 04:10UT to major bursts of bright tracks observed by HI near the elongation of \textit{PSP}, which can be traced back into the COR2A field of view. These episodes of large density increases, as measured by SPC, contain sequences of smaller density peaks separated by around 90-120 minutes. They are clearly reflected as additional narrower tracks in the J-map. They are shown in Figure~\ref{fig:JmapSTEREO}a and b by dashed vertical lines. Such density structures were noticed in past in situ measurements taken near 1 au \citep{Viall2008jgr}, in \textit{Helios} data between 0.3 and 0.6 au \citep{dimatteo2019} and separately in the spectral analysis of COR2A imagery \citep{Viall2015ApJ}. Combined \textit{STA} and \textit{PSP} observations allow, for the first time, investigation of the origin of individual features.\\

As a further check, we have analyzed the 3D trajectory of the density peaks using data from the full SECCHI field of view that extends out to 74$^\circ$ in elongation (corresponding to the outer edge of the HI2A field of view near the ecliptic plane). Over the time interval of interest here (April 6-12), most features completely fade as they near the outer edge of the HI1 field of view/inner edge of the HI2 field of view; this is expected for features that propagate in or near \textit{STA}'s plane of the sky. A trajectory analysis of these tracks using the fixed-phi technique (eq. \ref{eq:fixedphi}) that assumes plasma parcels move radially outwards at constant speed \citep[e.g.][]{Rouillard2008grl,Rouillard2009SoPh} yields values of $\beta$ ranging between 70$^\circ$ and 100$^\circ$, confirming that the features are expelled close to the plane of the sky during that time interval. \\



\newpage
\section{Discussion} \label{sec:Discussion}

The source regions of the different components of the slow solar wind are still debated. We know that the densest regions of the upper corona are associated with the bright coronal rays that emanate from helmet streamers. They have been long thought to generate the densest slow solar wind measured in situ, in particular hosting the source region of HPS that engulfs the HCS. Recent studies revisiting past data have argued that the slowest and densest solar wind measured in situ results from a magnetic connection in the vicinity of the coronal neutral line \citep{Sanchez2016JGRA}.\\ 

In this paper, we have combined multipoint imagery taken by \textit{STEREO} and \textit{SOHO} with the unprecedented in situ and remote-sensing observations made by \textit{PSP} of the nascent slow solar wind.

\begin{itemize}

    \item We make the first direct association between streamer rays and the dense solar wind measured in situ by \textit{PSP},
    
    \item We show that, as it moved to its southernmost heliographic latitudes near perihelion, \textit{PSP} briefly exited the streamer rays. It then entered a region that appeared darker than streamer rays in white-light images, precisely when SPC measured more tenuous and less variable plasma measured in situ.
    
    \item We demonstrate that \textit{PSP} remained on one side of the streamer belt around perihelion.
    
    \item We reveal a direct association between small white-light transients and density variations measured in situ by \textit{PSP} on timescales of tens of hours down to tens of minutes.
    
    \item We show that the white-light transients tracked to \textit{PSP} along the edge of the streamers contain many switchbacks associated with high densities.
    
\end{itemize}

These connections provide further context for interpreting the findings of \cite{Bale2019Nat} and \cite{Kasper2019Nat}. Because \textit{PSP} remained on one side of the streamer, the HCS was not measured during this period and remained in one polarity sector. \\

In addition, \textit{PSP} did not cross clear magnetic flux ropes that are expelled from the more central regions of the streamer. The last two decades of research have shown that streamer rays are continually perturbed by bursts of transient outflows \citep{sheeley97, Sheeley2010ApJ,Rouillard2011ApJ} released quasiperiodically from the top of helmet streamers. Multipoint imagery suggests they are formed by magnetic reconnection near 3-5\textit{R}$_\odot$ \citep{Sanchez2017ApJa,Sanchez2017ApJb}. These ``blobs'' normally disappear rapidly in the field of view of HI due to the drop in density associated with their expansion \citep[e.g.][]{Rouillard2008grl}. On occasion, these very slow transients get swept up by high-speed streams, thereby maintaining their high densities all the way to 1 au. The largest, and hence more massive, of these blobs have been tracked all the way to in situ spacecraft and have also been associated with the passage of helical magnetic fields \citep{Rouillard2010ApJ,Rouillard2010JGRA}. These flux ropes are measured in situ in the HCS \citep{Rouillard2009SoPh,Rouillard2011ApJ} between two sector boundaries.\\ 

\cite{Poirier2020ApJS} show that the zoomed-in view of streamer rays provided by WISPR enables a mapping of the small-scale morphology of streamers. This includes the densest part of the streamers where the HPS originates, this very high-density region that engulfs the HCS \citep{Winterhalter1994JGR}. We have used a WISPR Carrington map (Figure \ref{fig:carrmapW}) to compare the location of \textit{PSP} with this HPS at the start of the encounter (March 23-27); even then, \textit{PSP} remained several degrees south of the likely location of the HCS (orange arrows in Figure \ref{fig:carrmapW}). Individual WISPR images also show that flux rope structures were ejected continually northwards of \textit{PSP}. Therefore, the core of these flux rope structures were not expected to impact the \textit{PSP} spacecraft during that time interval. \\

Instead of a clean flux rope crossing, \textit{PSP} measured local reversals of the magnetic field direction that were not associated with 180$^\circ$ changes in the pitch angle of suprathermal electrons. These structures are interpreted in \cite{Bale2019Nat} and \cite{Kasper2019Nat} as folds in the magnetic fields that are often associated with density increases. They suggest that these higher densities should be detected in coronal and heliospheric imaging. The present study confirmes this, and connects the variable outflows from the streamers with strong bursts of ''switchbacks''. \\

Comparing Figure \ref{fig:contextinsitu} with Figure \ref{fig:carrmap}, we find evidence that these switchbacks have different properties inside and outside streamer flows (Figure~\ref{fig:contextinsitu}). The switchbacks in streamer flows tend to occur in bursts or clumps lasting several hours with sustained and significant changes in plasma density. Switchbacks originating from outside streamer stalks, likely from deeper inside coronal holes, are shorter lived and more numerous. These differences will be investigated further in an upcoming publication. \\

The present studies provides new clues to interpret the origin of the highly structured flows revealed by the analysis of deep-field \textit{STEREO}/COR2 observations \citep{DeForest2018}. The latter study predicted that \textit{PSP} would encounter  ``strong,  sharp variations in plasma density, by as much as an order of magnitude on timescales of 10 minutes or less.'' We have shown that the strong density variations are measured in situ by \textit{PSP} mainly inside streamer flows (Figure \ref{fig:contextinsitu}). We conclude that the strong density variations revealed by \textit{STEREO} are likely to originate inside and on the edges of streamer rays. The apparently ubiquitous nature of the density structures revealed by the analysis of \citet{DeForest2018} could be related to the presence of streamer rays at all PAs around the Sun. This would be expected at times of elevated solar activity that typically forces large excursions of the coronal neutral line and its associated helmet streamer. We also conclude that the structures imaged by \citet{DeForest2018} are likely to transport kinks and reversals in the magnetic field lines.  \\

Magnetic reconnection \citep[e.g.][]{Owens2018ApJ}, perhaps from chromospheric/coronal jets \citep[e.g.][]{Horbury2018}, and Kelvin Helmholtz instabilities \citep[e.g.][]{Suess2009} have been invoked as important physical mechanisms occurring at the interface between open and closed field lines, where the plasma escaping along open magnetic field lines meets the more static loop plasma. Both mechanisms could, in principle, produce folds in the magnetic field and plasma mixing at the boundary layers. \\

If we assume that switchbacks are formed by magnetic reconnection between open and closed magnetic field lines, a possible explanation for their different properties inside and outside streamer flows could reside in the size of the loops involved in the reconnection process. Streamer flows form above streamers where the largest coronal loops are typically adjacent to open magnetic field lines. These large loops are associated with dense plasma seen as bright helmet streamers in white-light images. In contrast, the smaller switchbacks measured just outside streamer flows could form in smaller loops lower in the corona. This idea could be tested in a future study by using composition data such as alpha to proton ratio changes associated with trains of density structures \citep{viall2009}. Future measurements of heavier ions by the \textit{Solar Orbiter} will be invaluable to investigate whether switchbacks inside and outside streamer flows contain different proportions of elements with low first-ionization potential \citep{laming2019}. \\

\section{Conclusion} \label{sec:Conclusion}

The \textit{PSP} mission is providing an unprecedented opportunity to connect solar winds with their source regions in the corona. This article has demonstrated the power of using multipoint and multi-instrument studies to study the sources of the slow solar wind. In doing so, we have made a clear connection between density variations  expelled along the edges of streamers and density structures measured in situ, providing new clues on the origin and structure of the slow solar wind. In future studies, we will attempt to link the magnetic properties of the small-scale transients with physical processes occurring in the corona using a combination of modelling and remote-sensing and in situ observations.\\ 

A better understanding of the dynamic outflows of streamers is important for other areas of solar physics. This region must host magnetic loop emergence and the periodic disconnection of open magnetic fields implicated in the long-term evolution of the open flux. The presence of switchbacks in the magnetic field was suggested in past studies of the total solar magnetic flux derived from in situ measurements \citep{Lockwood2009a}. Folds in the magnetic field were invoked as a source of the apparent increase of the total open flux (the “flux excess” effect) with heliocentric radial distance \citep{Lockwood2009b}. Recent studies have also found evidence that the highest energy particles could be accelerated when strong shocks reach the tip of streamers \citep[e.g.][]{Rouillard2016ApJ,Kouloumvakos2019ApJ}. The next decade of research with \textit{PSP} and the \textit{Solar Orbiter} promises to be rich in new discoveries on streamer flows. \\

\acknowledgments

The IRAP team acknowledges support from the French space agency (Centre National des Etudes Spatiales, CNES; \url{https://cnes.fr/fr}) that funds the plasma physics data center (Centre de Données de la Physique des Plasmas, CDPP; \url{http://www.cdpp.eu/}), the Multi Experiment Data \& Operation Center (MEDOC; \url{https://idoc.ias.u-psud.fr/MEDOC}) and the space weather team in Toulouse (Solar-Terrestrial Observations and Modelling Service, STORMS; \url{https://stormsweb.irap.omp.eu/}). This includes funding for the data mining tools AMDA (\url{http://amda.cdpp.eu/}) and CLWEB (\url{http://clweb.irap.omp.eu/}), and the propagation tool (\url{http://propagationtool.cdpp.eu}). A.K and Y.W. acknowledge financial support from the ANR project SLOW{\_}\,SOURCE (ANR-18-ERC1-0006-01), COROSHOCK (ANR-17-CE31-0006-01), and FP7 HELCATS project \url{https://www.helcats-fp7.eu/} under the FP7 EU contract number 606692. The work of A.P.R., N.P., L.G., V.R., P.L. was funded by the ERC SLOW{\_}\,SOURCE project (SLOW{\_}\,SOURCE - DLV-819189). The work of A.V., R.A.H., and G. S. was supported by the \textit{PSP}/WISPR project. N.M.V. is supported by the NASA Heliophysics Internal Scientist Funding Model. We thank the \textit{PSP}, \textit{STEREO}/SECCHI and \textit{SOHO}/LASCO teams.

%

\vspace{5mm}








\end{document}